\documentclass[a4paper,12pt]{article}

\usepackage[margin=0.75in]{geometry}

\usepackage{mathtools}  
\usepackage{mathrsfs}
\usepackage{amsmath}
\usepackage{amssymb}
\usepackage[margin=0.75in]{geometry}
\usepackage{textcomp}
\usepackage{array}
\usepackage[usenames, dvipsnames]{color}
\usepackage[font=small]{caption}
\usepackage{floatrow}
\usepackage{subfig}
\usepackage{cite}
\usepackage[utf8]{inputenc}
\usepackage{amsthm}
\usepackage{multicol}

\usepackage{hyperref}
\newcommand\blfootnote[1]{%
  \begin{NoHyper}%
  \renewcommand\thefootnote{}\footnote{#1}%
  \addtocounter{footnote}{-1}%
  \end{NoHyper}%
}

\numberwithin{equation}{section}

\newcolumntype{C}{>{$}c<{$}} 		
\newcolumntype{L}{>{$}l<{$}} 		

\newcommand{\kahl}{K\"ahler}
\newcommand{\w}{\ensuremath{\wedge}}

\newcommand{\mc}{\mathcal}
\newcommand{\mbb}{\mathbb}

\newcommand{\be}{\begin{equation}}
\newcommand{\ee}{\end{equation}}
\newcommand*{\nnbe}{\begin{equation}}
\newcommand*{\nnee}{\end{equation}}
\newcommand{\bea}{\begin{eqnarray}}
\newcommand{\eea}{\end{eqnarray}}
\newcommand{\ba}{\begin{align}}
\newcommand{\ea}{\end{align}}
\newcommand{\bi}{\begin{itemize}}
\newcommand{\ei}{\end{itemize}}

\newsavebox{\overlongequation}

\newcommand{\fref}[1]{Figure~\ref{#1}}
 
\newcommand{\fthb}{B_3}				
\newcommand{\cand}[1]{D_{K_{#1}}}		
\newcommand{\canb}[1]{K_{#1}}			
\newcommand{\candl}{\overline{D}_{K_{B_3}}}	

\newcommand{\cyu}{X}					
\newcommand{\cyo}{X_o}				
\newcommand{\ori}{\sigma}				

\newcommand{\as}{\mc{A}}				
\newcommand{\asl}{\overline{\mc{A}}}	

\newcommand{\db}{S}					
\newcommand{\dbl}{\overline{S}}		

\newcommand{\crv}{\gamma}			
\newcommand{\crvl}{\overline{\gamma}}	
\newcommand{\degr}[1]{\tilde{d}_{#1}}			

\newcommand{\flxcont}[1]{Q_{#1}}		
\newcommand{\stbmod}[1]{n_{\mathrm{D7},{#1}}^\mathrm{stab.}}		

\newcommand{\tanb}[1]{T_{#1}}			
\newcommand{\nrmb}[2]{\mc{N}_{{#1}\backslash{#2}}}

\newcommand{\ec}{\chi}				
\newcommand{\ind}{\mathrm{ind}}		

\begin{document}

\begin{titlepage}
\begin{center}
\rightline{\small }

	\vskip 2cm
	{\Large \bf D7 Moduli Stabilization: The Tadpole Menace}
	
	\vskip .6cm

	{ Iosif Bena$^1$, Callum Brodie$^{1,2}$, and Mariana Gra\~na$^1$}
	
	\vskip 1cm
      
         ${}^1$ {\it 
	Institut de Physique Th\'eorique,
	Universit\'e Paris Saclay, CEA, CNRS\\
	Orme des Merisiers, 91191 Gif-sur-Yvette Cedex, France
	}
	
	\vskip 0.2cm
       
       ${}^2$ {\it 
	Department of Physics, Robeson Hall, Virginia Tech \\
	Blacksburg, VA 24061, U.S.A.
	}
\end{center}

\vskip 1cm

\begin{center} {\bf Abstract }\\
\end{center}

\noindent D7-brane moduli are stabilized by worldvolume fluxes, which contribute to the D3-brane tadpole. We calculate this contribution in the Type IIB limit of F-theory compactifications on Calabi-Yau four-folds with a weak Fano base, and are able to prove a no-go theorem for vast swathes of the landscape of compactifications. When the genus of the curve dual to the D7 worldvolume fluxes is fixed and the number of moduli grows, we find that the D3 charge sourced by the fluxes grows faster than 7/16 of the number of moduli, which supports the Tadpole Conjecture of Ref.~\cite{Bena:2020xrh}. Our lower bound for the induced D3 charge decreases when the genus of the curves dual to the stabilizing fluxes increase, and does not allow to rule out a sliver of flux configurations dual to high-genus high-degree curves. However, we argue that most of these fluxes have very high curvature, which is likely to be above the string scale except on extremely large (and experimentally ruled out) compactification manifolds.

\blfootnote{iosif.bena@ipht.fr}
\blfootnote{callumb@vt.edu}
\blfootnote{mariana.grana@ipht.fr}

\vfill

\end{titlepage}

\tableofcontents
\pagebreak

\section{Introduction}

Compactifications of String Theory to four dimensions come with a very large number of moduli - massless scalar fields that must be given a mass if the resulting theory is to have anything to do with the real world. Thus, the stabilization of moduli and its interplay with supersymmetry-breaking is perhaps the most important open problem of String Phenomenology.

String compactification moduli come in different flavors, which change as one moves to different duality frames. In particular, Type IIB compactifications on Calabi-Yau manifolds have complex structure and K\"ahler moduli, that correspond to deformations of the compactification manifold, as well as D3, D5 and D7 moduli that correspond to deformations of the branes that are wrapped on this manifold. If one describes this compactification in the language of F-theory, both the complex structure moduli and the D7 moduli appear as complex structure moduli of the F-theory four-fold, despite their seemingly different IIB origin.

Stabilizing moduli comes at a cost. Type IIB complex structure moduli are stabilized by turning on fluxes on the compactification manifold, which give a non-trivial contribution to the D3-brane tadpole. Similarly, D7-brane moduli are stabilized by turning on certain holomorphic worldvolume fluxes, which again contribute to the D3-brane tadpole. On the other hand, K\"ahler moduli are stabilized by non-perturbative effects: instantons or D7-brane gaugino condensation \cite{Kachru:2003aw}, which can only be treated at the level of effective field theory. 

In \cite{Bena:2020xrh,Bena:2021wyr}, J.~Bl{\aa}b{\"a}ck, S.~L\"ust and two of the authors have argued that both complex structure moduli stabilization and D7 moduli stabilization are plagued by the Tadpole Problem: The fluxes that stabilize the moduli source a positive D3 charge $Q_\mathrm{D3}$ whose minimal value is conjectured to grow linearly with the number of moduli
\be
Q_{\rm D3}^{\rm stab} \ge \alpha \, n_{\rm moduli}\, \nonumber .
\label{tadpole}
\ee
Furthermore, based on known examples, they have conjectured that in the large $n_{\rm moduli}$ limit, the linearity coefficient, $\alpha$, is larger than $1/3$, which is above the upper bound allowed by tadpole cancelation.

Proving the Tadpole Conjecture would rule out all String Theory compactifications with large numbers of stabilized D7 or complex structure moduli, and also rule out de Sitter vacua obtained by uplifting anti de Sitter compactifications using antibranes in warped throats \cite{Kachru:2003aw}, since one needs a large $Q_{\rm D3}$ to avoid instabilities \cite{Bena:2018fqc}.  The purpose of this paper is to offer the next best thing to a proof: a calculation of the charge induced by the fluxes needed to stabilize the D7 moduli in a huge  family of compactifications, which spectacularly confirms the linear growth of the D3 charge with the number of moduli that are stabilized. 

We consider F-theory compactifications on CY four-folds that are elliptic fibrations over a three-fold base, $\fthb$. The stabilization of D7 moduli by worldvolume flux for $\fthb \cong \mbb{P}^3$ was analyzed in Ref.~\cite{Collinucci:2008pf}. To preserve supersymmetry, the flux should be a holomorphic and anti-self-dual (and hence a primitive) (1,1) form on the four-dimensional compact cycle wrapped by the D7-branes. The key ingredient in stabilizing D7 moduli is to realize that there exist special loci in the configuration space of these divisors where new (1,1) forms, dual to holomorphic curves, appear on the divisor; by turning on worldvolume flux components along these new directions, the D7-brane ends up being supersymmetric only when wrapping these special divisors, and supersymmetry is broken when the D7-brane moves away. This stabilizes the D7-brane to the special loci where the new (1,1) forms appear. By calculating the D3 charge tadpole sourced by these (1,1) fluxes, \cite{Collinucci:2008pf} found that D7 moduli cannot be stabilized within the tadpole bound. From their calculation we can furthermore conclude that the induced D3 charge grows linearly with the number of moduli, and the ratio between them is larger than $1/3$, thus verifying the Tadpole Conjecture. 

One straightforward generalization of the methods in Ref.~\cite{Collinucci:2008pf} is to base spaces which are toric and weak Fano. In this case the flux needed to stabilize D7 moduli can be computed using the help of toric geometry methods, and we find again a linear relation between number of D7 moduli and the charges induced by the fluxes needed to stabilize these moduli.

In this paper we attack the more complicated problem of base spaces that are weak Fano, but not necessarily toric. Additionally, while in Ref.~\cite{Collinucci:2008pf} the (1,1) flux was restricted to be dual to a curve of genus zero to allow an explicit description of the flux curve embedding, we consider the case of arbitrary genus. We compute the positive D3 charge contribution of the (1,1) flux that stabilizes the D7 moduli for a given degree and genus of the curve describing this flux, and show that the D3 charge has a lower bound that grows linearly with the degree associated to this curve. Next we relate the number of moduli stabilized by the flux to the degree of the curve dual to the flux, and argue that this number has an upper bound, which also grows linearly with the degree of this curve! As one can see in Appendix~\ref{app:tadpole} and Appendix~\ref{app:mod_count} where these calculations are fleshed out in detail, there are several steps of the computation that involve quantities that are not generic and that are hard to calculate without specific knowledge of the details of the base space and the flux curve embedding. However, one can show that these hard-to-compute terms always have a definite sign, and make the D3 charge larger than our bound.

On the one hand, our results spectacularly confirm the linear growth of the D3 charge sourced by D7 fluxes with the number of moduli these fluxes stabilize. Indeed, if one fixes the genus of the curve dual to the D7 worldvolume flux, both the tadpole and the number of stabilized moduli grow linearly with the degree of the curve. This proves the tadpole conjecture in this context for D7 stabilization by flux curves of constant genus and growing degree. Furthermore, we find that the tadpole conjecture proportionality constant, $\alpha$, is equal to $\frac{7}{16} = 0.4375$, a value that is very close to the values found in the four examples discussed in Ref.~\cite{Bena:2020xrh}.

On the other hand, one can also attempt to stabilize D7 moduli by curves whose degree and genus grow together. Since the genus of the curve lowers our bounds on the induced D3 charge without affecting the number of moduli stabilized, our calculation does not rule out the possibility that, for a sufficiently large genus, one could stabilize all the D7 moduli and evade the Tadpole Conjecture. The region which is not ruled out by our calculation is a narrow sliver in the degree-genus plane, depicted in Figure~\ref{fig:constraint_plot}. We argue however that D7-brane fluxes in this sliver have such large genera that it is likely that their curvature can only be below string-scale when the size of the compactification manifold is very large. We discuss this in detail in Section~\ref{sec:window}.

The paper is organized as follows: In Section~\ref{sec:setup} we discuss the problem of D7 moduli stabilization by worldvolume flux. In Section~\ref{sec:nogo_review} we review the example in Ref.~\cite{Collinucci:2008pf}, where it was shown that the D7 moduli cannot be stabilized within the tadpole bound. In Section~\ref{sec:gen} we analyze moduli stabilization for F-theory compactifications with a weak Fano base. Appendices~\ref{app:geom}, \ref{app:tadpole}, and \ref{app:mod_count} contain the details of the calculation for general weak Fano bases, while Appendix~\ref{app:toric} presents a more explicit calculation of the tadpole when the base is toric.

\section{Stabilizing D7-brane moduli}
\label{sec:setup}

\subsection{D7-branes in Type IIB orientifolds}
\label{sec:setup_geom}

In this section we briefly recall the relation between F-theory and Type IIB orientifolds, which will be heavily used in the following sections.

In the F-theory framework, the background geometry is a four-fold $Z_4$, which in particular is an elliptic fibration: a torus-fibration with a section, over a three-fold base, $\fthb$. We assume the four-fold is smooth and has strict SU(4) holonomy\footnote{If the holonomy group were a subgroup of SU(4), the background would have more supersymmetry and the moduli stabilization mechanism would work in a slightly different way.}. The elliptic curve over each point of the base $\fthb$ can be described as a hypersurface inside the weighted projective space $\mbb{P}^2_{231}[u:v:w]$, as
\be
v^2 = u^3 + f u w^4 + g w^6 ,
\ee
where the coefficients $f$ and $g$ control the geometry of the elliptic curve. The variation of the coefficients $f$ and $g$ over the base $\fthb$ fixes the geometry of the four-fold $Z_4$ for a given base space.

We will illustrate all the concepts in the simple example when the base space is the complex projective space $\mbb{P}^3(x)$, so that $f(x)$ and $g(x)$ vary over $\fthb \cong \mbb{P}^3(x)$ as homogeneous polynomials in $x$ of degrees 16 and 24 respectively\footnote{More generally, $f$ and $g$ are sections $f \in \Gamma\big((\canb{\fthb}^*)^{\otimes 4}\big)$ and $g \in \Gamma\big((\canb{\fthb}^*)^{\otimes 6}\big)$ of powers of the anti-canonical bundle $\canb{\fthb}^*$ of the base $\fthb$. We also note that the coordinates on the $\mbb{P}^2_{231}$ are sections $u \in \Gamma\big((\canb{\fthb}^*)^{\otimes 2}\big)$, $v \in \Gamma\big((\canb{\fthb}^*)^{\otimes 3}\big)$, and $w \in \Gamma\big((\canb{\fthb}^*)^{\otimes 1}\big)$.}.

From the Type IIB perspective, the above data define a compactification of perturbative IIB string theory on the base three-fold $\fthb$ with D7-branes and O7-planes. The complex structure of the elliptic curve in $Z_4$, which varies over $\fthb$, plays the role  of the axio-dilaton field $\tau = C + e^{-\phi}$.

For general choices of $f$ and $g$, the profile of $\tau$ gives rise to large string couplings (and hence no perturbative Type IIB description) over much of $\fthb$. However, for any given $f$ and $g$ there exists a limiting deformation, the Sen limit \cite{Sen-limit}, which gives a Type IIB weakly-coupled description of all of $\fthb$ except over a lower dimensional subset. In particular, parameterizing the functions $f$ and $g$ without loss of generality as
\be
f = -3h^2 + \epsilon \eta \,, \quad g = -2h^3 + \epsilon h \eta - \tfrac{1}{12}\epsilon^2 \chi \,,
\ee
where $\epsilon$ is a constant, and where it follows that $h$, $\eta$, and $\chi$ vary over the base $\fthb$ as homogeneous polynomials\footnote{For a more general base $\fthb \ncong \mbb{P}^3$, these are sections $h \in \Gamma\big((\canb{\fthb}^*)^{\otimes 2}\big)$, $\eta \in \Gamma\big((\canb{\fthb}^*)^{\otimes 4}\big)$, and $\chi \in \Gamma\big((\canb{\fthb}^*)^{\otimes 6}\big)$.} of degree 8, 16, and 24 respectively, one finds that in the limit $\epsilon \to 0$ the string coupling goes to zero almost everywhere, except at codimension-one loci, where a monodromy analysis reveals the presence of D7-branes and O7-planes, specifically where the following equations are satisfied
\be
\mathrm{O7}:~h(x)=0 \,, \quad \mathrm{D7}:~\eta^2(x)=h(x)\chi(x) \,.
\label{eq:o7_and_d7_loci}
\ee
The fact that the O7-plane locus is at $h(x)=0$ means that the double-cover of the orientifold $\cyo \cong \fthb$ can be described as\footnote{Generically, if the D7-branes are not on top of the O7-planes, the variation of the axio-dilaton is such that the supersymmetric solution requires the metric on $X$ to be non-Ricci flat \cite{Grana:2001xn}. Nevertheless, since the double-cover of the orientifold admits a Ricci-flat metric, we will refer to it as a CY manifold, or CY orientifold.} 
\be
\cyu: ~ \xi^2 = h(x) \,,
\label{eq:cy_eqn}
\ee
which introduces an additional complex coordinate, $\xi$, to describe $\cyu$ as a hypersurface inside an ambient space, $\as$. When the base space is $\fthb \cong \mbb{P}^3[x]$, this ambient space, $\as$, is the weighted projective space $\mbb{P}^4[\xi:x]$, where the coordinates $\xi$ and $x_i$ ($i=0$ to $3$) satisfy the equivalence $[\xi:x] \sim [\lambda^4 \xi: \lambda x]$ for $\lambda \in \mbb{C}^*$. Note that identifying points on $\cyu$ under the orbifold action $\ori: \xi \to -\xi$ gives back $\fthb$. 

\bigskip

The above descriptions of the CY and the D7-brane content allow for a straightforward computation of the number of D7-brane deformation moduli. The D7-brane locus is given by the intersection $\{\xi^2 = h(x) \} \cap \{ \eta^2(x)=h(x)\chi(x) \}$ inside the ambient space $\as$. Furthermore, the function $h(x)$ is fixed by the specification of the CY, and the configuration of the D7-brane is specified by the functions $\eta(x)$ and $\chi(x)$. However, the equation for the D7-brane is invariant under the simultaneous shifts
\be
\eta \to \eta + h \psi \,, \quad \chi \to \chi + 2 \eta \psi + h \psi^2 \,,
\ee
where $\psi$ is arbitrary, and this redundancy must be subtracted. Additionally, since the D7-brane is specified only by the zero locus of the equation, there is also one irrelevant overall scaling.

In the simple example of a base $\fthb \cong \mbb{P}^3[x]$, the functions $\eta$, $\chi$, and $\psi$ are homogeneous polynomials of degree 16, 24, and 8 in the coordinates $x_i$, and the number of D7-brane deformation moduli is given by
\be
n_{\mathrm{D}7} (\mbb{P}^3)=\binom{16+3}{3}+\binom{24+3}{3}-\binom{8+3}{3}-1 = 3728 \,.
\ee
For a general base space, the parameters in $\eta$, $\chi$, and $\psi$ are counted by the dimensions of the spaces of sections of the relevant bundles and the number of D7-brane deformation moduli is 
\be
n_{\mathrm{D}7}=h^0\big(\fthb,\,(\canb{\fthb}^*)^{\otimes 4}\big) + h^0\big(\fthb,\,(\canb{\fthb}^*)^{\otimes 6}\big) - h^0\big(\fthb,\,(\canb{\fthb}^*)^{\otimes 2}\big) - 1 \,.
\label{eq:num_d7_mod_gen}
\ee
We note that when the base, $\fthb$, is weak Fano (which is a condition we will return to in Section~\ref{sec:gen}) one can reduce this sum to the compact expression
\be \label{eq:nD7}
n_{\mathrm{D}7}=16 + 58\int_{\fthb}c_1(\fthb)^3 \,,
\ee
so that the number of D7-brane deformation moduli is written purely in terms of a Chern number of the base space, $\fthb$. We give the details of this computation in Appendix~\ref{app:geom}.

\bigskip

The F-theory description also provides a simple way to compute the D3 tadpole. In particular, all of the negative-charge contributions to the tadpole coming from O7-planes and D7-branes are captured in the topology of the four-fold $Z_4$, specifically by its Euler number, $\ec(Z_4)$, via
\be \label{eq:tad_neg_cont}
Q_\mathrm{neg.} = -\,\frac{\ec(Z_4)}{24} \,.
\ee
This topological quantity gives an upper bound on the positive-charge contributions, which come from 3-form flux, D7-brane worldvolume flux, and mobile D3-branes.

In the simple example of a base $\fthb \cong \mbb{P}^3[x]$, this D3 charge bound is given by $\frac{\ec(Z_4)}{24} = \frac{23328}{24} = 972$. For a generic base space one can show that (see Appendix~\ref{app:geom})
\be
\frac{\ec(Z_4)}{24} = 12 + 15\int_{\fthb}c_1(\fthb)^3  \,.
\label{eq:tadp_gen}
\ee
Again this expression involves a single (and the same) Chern number as in Equation~\eqref{eq:nD7}.

From Equations \eqref{eq:nD7} and \eqref{eq:tadp_gen} we can see that the ratio between the number of D7 moduli and the negative D3 tadpole contribution of the D7-branes in the limit of a large number of moduli is
\be \label{tadpolenD7}
|Q_\mathrm{neg.}| \sim \frac{15}{58} \, n_{\mathrm{D}7} \ .
\ee
The proportionality constant $15/58=0.259$ is slightly greater than $1/4$, the value found in Ref.~\cite{Bena:2020xrh} by examining general F-theory compactifications.\footnote{This discrepancy likely arises because the limit of large $h^{3,1}$  \cite{Bena:2020xrh} is not the same as the large $c_1^3$ limit.}
Hence, if the charge sourced by the fluxes needed to stabilize the $n_{\mathrm{D}7}$ moduli is larger than $\frac{15}{58}\,n_{\mathrm{D}7}$, it will be impossible to stabilize all the D7 moduli within the tadpole bound.

\subsection{Stabilizing D7-brane moduli with worldvolume flux}
\label{sec:setup_stab}

A D7-brane carries a worldvolume flux, which is specified by a two-form, $F$. For simplicity we consider throughout a single irreducible D7-brane, for which the worldvolume flux is that of a U(1) gauge theory.

\bigskip

For the worldvolume flux to preserve supersymmetry, the two-form $F$ must be anti-self-dual,\ $F = -*F$, where $*$ is the Hodge star on the D7-brane. This condition turns out to be equivalent to imposing the two constraints \cite{Marino:1999af}
\be
F^{0,2} = F^{2,0} = 0 \,, \quad F \w J = 0 \,,
\label{eq:sups_conds}
\ee
where $J$ is the {\kahl} form of the orientifold pulled back to the D7-brane, and the first condition means that $F$ is a form of type (1,1) with respect to the complex structure on the D7-brane.

Since the complex structure on the D7-brane is inherited from that of the CY orientifold, $\cyo$, in which it is embedded, deformations of the D7-brane alter its complex structure. A particularly important consequence is that for a subset of configurations, there may appear new holomorphic (1,1)-forms on the D7-brane worldvolume, which are hence new candidates for the flux $F$.

By tuning the D7-brane to be in such a special locus in the moduli space and by turning on flux corresponding to a newly appearing (1,1)-form, the D7-brane configuration is energetically fixed within this locus, since deformations away from this locus necessarily render the flux non-supersymmetric. Hence, this worldvolume flux stabilizes the D7-brane deformation moduli.

\bigskip

Since $F$ is a form of type (1,1), it is dual to a linear combination of holomorphic curves. From this point of view, stabilization with worldvolume flux is possible when there are holomorphic curves in the CY which are not generically contained in the D7-brane, but which are contained by particular configurations of the D7-brane. We note that since the expected dimension of the space of available deformations of a curve in a CY three-fold is zero\footnote{In particular, the Euler characteristic $\chi(\nrmb{C}{X}) = h^0(C,\nrmb{C}{X}) - h^1(C,\nrmb{C}{X})$ of the normal bundle of the curve is zero. This follows from the defining exact sequence $0 \to \tanb{C} \to \tanb{X}|_C \to \nrmb{C}{X} \to 0$ of the normal bundle, and the fact that $\chi(\tanb{X}|_C) = 3-3g$ and $\chi(\tanb{C}) = 3-3g$. The deformations of $C$, which correspond to $H^0(C,\nrmb{C}{X})$, are then generically totally obstructed by $H^1(C,\nrmb{C}{X})$.}, one can in general expect the existence of many isolated curves, which are useful candidates for this stabilization mechanism.

In addition to the anti-self-dual condition, the flux configuration must be consistent with the orientifold projection. This requires that $F = -\ori^*F$, where $\ori$ is the orbifold action induced on the D7-brane. In the geometric picture of the flux, oddness under the orbifold action means that the flux must be described by a sum of curves of the form
\be
F \, \sim \, \crv - \crv' \,,
\label{eq:flux_form}
\ee
where $\crv' = \ori \crv$ is the image of the curve $\crv$ under the orbifold map.

We note that this combination of fluxes automatically satisfies the second condition $F \wedge J = 0$ in Equation~\eqref{eq:sups_conds} at the level of the cohomology classes, since the {\kahl} form, $J$, is even under the orbifold action while the above flux (class) $F$ is odd. However, this is not sufficient to impose the condition at the level of the forms.\footnote{In Ref.~\cite{Collinucci:2008pf}, below Equation~(3.59), it is said that this choice of flux class ensures the existence of representatives that satisfy the full condition $F \wedge J = 0$. However in general it is not true that any exact piece in $F \wedge J$ can be cancelled by exact additions to $F$ and $J$. In Ref.~\cite{Collinucci:2008pf} the statement is based on the claim that $F \wedge J$ is harmonic if $F$ is harmonic, but this is not true in general. We are not aware of any other argument for the existence of appropriate representatives, and indeed in Section~\ref{sec:gen}, we will need to rule out by hand some fluxes with classes of the form \eqref{eq:flux_form} which give a negative contribution to the tadpole and hence cannot satisfy $F \wedge J = 0$.}

Indeed at the level of our analysis, which is based on topology and which avoids detailed calculations of the geometry, we are not aware of a straightforward way to impose the genuine vanishing of $F \wedge J$. Hence, in the analysis of Sections~\ref{sec:nogo_review} and \ref{sec:gen}, we will content ourselves with imposing only this weaker condition on the classes. Notably, this means that there may exist a significant strengthening of the constraints we derive in Section~\ref{sec:gen} on the set of possible stabilizing flux configurations.

\bigskip

A worldvolume flux on the D7-brane contributes to the D3 tadpole:
\be
Q_F = - \frac{1}{4}\int F \w F \,. \label{D3tad}
\ee
This contribution is manifestly positive after imposing $F = -* F$. The computation of this quantity is subtle when the D7-brane surface $\db$ is singular, but can be performed by utilizing a resolved version $\dbl$ of the D7-brane. This will be discussed in detail in Appendix~\ref{app:tadpole}. In the geometric picture of a curve dual to the flux, the integral that gives the D3 charge sourced by the flux \eqref{D3tad} becomes a curve intersection, and for the $\ori$-odd form in Equation~\eqref{eq:flux_form} it is
\be
\flxcont{\crv} = - \frac{1}{4}(\crvl - \crvl') \cdot (\crvl - \crvl') = \frac{1}{2}\big(-\crvl^2 + \crvl \cdot \crvl'\big) \,,
\label{eq:tadp_cont_gen}
\ee
where $\crvl$ is the lift of the curve $\crv$ to the resolved version, $\dbl$, of the D7-brane surface, and where the curve intersections are computed on $\dbl$. Below we will be interested in how the size of this contribution grows as the flux becomes sufficient to completely stabilize the D7-brane deformation moduli.\footnote{It is important to note that because of the Freed-Witten anomaly the flux is not generically an element of $H^2(\dbl,\mbb{Z})$ but rather an element of the shifted lattice $H^2(\dbl,\mbb{Z}) + \frac{1}{2}c_1(\dbl,\mbb{Z})$. Since we are mostly interested the large  $c_1(\fthb)^3$ limit, we expect that this is a small effect, which we hence neglect.}

\section{Review: a no-go example}
\label{sec:nogo_review}

In Ref.~\cite{Collinucci:2008pf}, the authors showed that when the F-theory base space, $\fthb$, is the complex projective space $\mbb{P}^3[x]$, one cannot stabilize all D7-brane moduli using worldvolume flux. Here we review their argument, before turning to generalizations in Section~\ref{sec:gen}.

The compactification space and the  D7-brane worldvolume flux considered in Ref.~\cite{Collinucci:2008pf} are taken to be possibly the simplest available choices: the F-theory four-fold is taken to be a smooth elliptic fibration over the projective space $\mbb{P}^3$, and the stabilizing flux discussed in Section~\ref{sec:setup_stab} is restricted to be dual to a curve of genus zero (a sphere $\mbb{P}^1$).

\bigskip

Because both the base space, $\fthb = \mbb{P}^3[x]$, and the embedded curve, $\crv = \mbb{P}^1[u]$, are simply parameterized by coordinates $x_0,\,x_1,\,x_2,\,x_3$ and $u_0,\,u_1$, it is possible to describe the embedding $\crv \xhookrightarrow{} \cyu$ of the curve into the CY three-fold with a polynomial map. This has the notable benefit of allowing for a straightforward computation of the number of D7-brane moduli stabilized by this flux.

As the base $\fthb$ of the F-theory four-fold is $\mbb{P}^3$, the ambient space, $\as$, of the double-cover of the Type IIB orientifold is also simply parameterized by coordinates $\xi,\,x_0,\,x_1,\,x_2,\,x_3$, and hence we can specify the embedding $[u_0 : u_1] \to [\xi : x]$ of the curve $\crv$ into the ambient space $\as$ as
\be
\crv \, : \quad \xi = \Xi[u_0,u_1] \,, ~ x_i = X_i[u_0,u_1] \,,
\label{eq:crv_into_as}
\ee
where $\Xi$ and $X_i$ are respectively degree $4d$ and degree $d$ homogeneous polynomials in $u_0$ and $u_1$. Here the factor of 4 is required for compatibility with the projective identification on $\as$, namely \ $[\xi:x] \sim [\lambda^4 \xi: \lambda x]$ for $\lambda \in \mbb{C}^*$.

For this polynomial map to embed the curve into the upstairs CY, $\cyu$, the image of the embedding, must satisfy the defining equation of the CY for every point on the curve,
\be
\Xi^2(u_0,u_1) - h\big(X(u_0,u_1)\big)\stackrel{!}{=}0 ~~ \forall \;[y_0:y_1]\,.
\ee
This equation is of degree $8d$ in $u_0$ and $u_1$, and hence contains $8d+1$ coefficients, which must all be zero for the polynomial to vanish identically. But the number of free parameters in the embedding map is $(4d+1)+4(d+1) - 4 = 8d+1$, where the two bracketed terms count the coefficients in $\Xi$ and the $X_i$ while the term $-4$ comes from subtracting the $\mathrm{GL}_2$ reparameterizations of the $\mbb{P}^1[u]$. Hence, for any degree $d$, there are polynomial embeddings as in \eqref{eq:crv_into_as} which indeed determine a curve inside the CY, and these curves do not form a continuous family but are instead isolated. These are hence rigid curves which provide ideal candidates to stabilize the D7-brane moduli, as described in Section~\ref{sec:setup_stab}.

To use the above curves for stabilization, we demand that they be contained inside the D7-brane. Hence the image of the embedding must satisfy the equation for the D7-brane:
\be
\eta^2\big(X(u_0,u_1)\big) = \Xi^2(u_0,u_1)\,\chi\big(X(u_0,u_1)\big) ~~ \forall \;[y_0:y_1]\,.
\ee
Since the curves are rigid, this equation directly provides constraints on the D7-brane configuration. In particular, this equation is of degree $32d$, and hence provides $32d+1$ constraints.
In order to completely stabilize the D7-brane, this number of constraints must exceed the number of independent deformations of the D7-brane. In the example we discuss, we have shown in Section~\ref{sec:setup_geom} that the number of independent deformations is 3728. Hence, comparing to $32d+1$, we see that the complete stabilization of the D7-brane requires that the degree of the curve satisfies
\be
32d+1\ge 3728 \quad \Rightarrow  \quad d  \geq 117 \,.
\label{eq:mod_stab_p3}
\ee

\bigskip

The flux dual to a curve of such a high degree will have a large contribution to the D3 tadpole. This contribution can be written in terms of curve intersections as (c.f. Eq. \eqref{eq:tadp_cont_gen}),
\be
\flxcont{\crv} = \frac{1}{2}\big(-\crvl^2 + \crvl \cdot \crvl'\big) \,,
\ee
where $\crvl$ is the lift of the curve $\crv$ to the resolved version, $\dbl$, of the D7-brane, and $\crvl'$ is its image under the orbifold map, and where the curve intersections are computed on $\dbl$. The first term can be rewritten using the relation
\be
-\crvl^2 = \chi(\crvl)-\crvl \cdot c_1(\tanb{\dbl}) \,.
\ee
Since $\crvl$ is genus zero, the Euler characteristic is simply $\chi(\crvl) = 2-2g = 2$. Computing  the term $\crvl \cdot c_1(\tanb{\dbl})$ requires knowledge of the geometry of the resolved D7-brane, $\dbl$, which we relegate to Appendix~\ref{app:tadpole}. We find, as in Ref.~\cite{Collinucci:2008pf}, that for a curve $\crv$ of degree $d$,
\be
-\crvl^2 = 2 + 28d \,.
\ee
Since the second term $\crvl \cdot \crvl'$ in $\flxcont{\crv}$ is an intersection between distinct irreducible curves, it is non-negative by definition\footnote{In fact one can argue that this intersection is generically zero, but it is sufficient for our purposes that it is always non-negative.}, and hence the above term provides a lower bound,
\be
\flxcont{\crv} \geq 1 + 14d\,. 
\label{eq:tadp_cont_simpl_ex}
\ee
Hence, for a curve $\crv$ of degree $d \geq 117$, the contribution to the D3 tadpole is at least
\be
\flxcont{\crv} \geq 1638 \,.
\ee
Notably, this is significantly larger than the allowed maximum value given in Equation~\eqref{eq:tad_neg_cont}, which for the present four-fold, with a $\mbb{P}^3[x]$ base, is
\be
|Q_\mathrm{neg.}| = 972 \, . 
\ee
Hence, the authors of Ref.~\cite{Collinucci:2008pf} were able to conclude that in this compactification it is not possible to stabilize all D7-brane moduli with worldvolume flux without overshooting the D3 tadpole bounds.

\section{A general compactification with a weak Fano base}
\label{sec:gen}

The moduli stabilization scenario worked out in Ref.~\cite{Collinucci:2008pf} utilized perhaps the simplest F-theory base space, $\mbb{P}^3$, and a stabilizing D7 worldvolume flux dual only to a holomorphic curve of genus zero. The main question we want to answer in this paper is whether the relation between the number of stabilized moduli and the D3 tadpole, and the impossibility to stabilize moduli within the tadpole bound, are artifacts of the simplicity of the construction in Ref.~\cite{Collinucci:2008pf}, or rather are generic features of D7 moduli stabilization.

There are several possible ways to enlarge the families of compactifications. One particularly straightforward construction is to take the F-theory base space to be any smooth toric weak Fano three-fold. Here toric geometry methods allow us to describe the stabilization of the D7-brane very explicitly, making the count of the number of stabilized moduli straightforward. The details of these calculations are given in Appendix~\ref{app:toric}.\footnote{The results we obtain for toric weak Fano bases are in line with those we obtain for general weak Fano bases, which we discuss in this section. The no-go theorem we deduce in Appendix~\ref{app:toric} is a special example of the more general no-go theorem we present here.} We note that this result already applies to a very large class of compactifications, as the number of toric weak Fano three-folds is estimated to be $O(10^{15})$ \cite{Halverson_2017}.

However, one can do even more.  First, we can allow the stabilizing worldvolume flux to be dual to a holomorphic curve of arbitrary genus, rather than only genus-zero curves. Second, we can take the F-theory base space to be any smooth weak Fano three-fold (not necessarily toric). Amazingly, despite the very general nature of the D7-branes and of their stabilizing fluxes, one can still constrain the relation between flux-stabilized moduli and flux-induced tadpole, as we will see below.

The primary difficulty in attacking this very general problem is to compute the number of stabilized D7-brane moduli. Following Section~\ref{sec:setup_stab}, we recall that the mechanism to stabilize the D7-brane consists of tuning the brane to contain a particular holomorphic curve, $\crv$, of arbitrary genus $g_\crv$, and turning on worldvolume flux dual to $\crv$. The extra difficulty in the case of arbitrary genus is that the embedding of the curve $\crv$ can no longer be specified explicitly, so that one requires more abstract tools from algebraic geometry.

\bigskip

The D3 charge contribution, $\flxcont{\crv}$, of the worldvolume flux dual to a curve $\crv$ can be computed by resolving the singular D7-brane (as discussed in Section~\ref{sec:setup_stab}) and computing the flux curve intersection in Equation~\eqref{eq:tadp_cont_gen} of the resolved D7-brane. We relegate the details of this computation to Appendix~\ref{app:tadpole}. As for the genus-zero flux curve in Section~\ref{sec:nogo_review}, the term $\crvl \cdot \crvl'$ can generically be expected to be zero. However, if present, it can only increase the D3 tadpole. One thus finds a lower bound for the D3 tadpole contribution, $\flxcont{\crv}$, of the flux dual to a curve with genus $g_\crv$:
\be
\flxcont{\crv} \geq (1-g_\crv) + \tfrac72 \degr{\crv} \,,
\label{eq:tadp_contr_gen}
\ee
where we have defined the canonical degree, $\degr{\crv}$, of the curve $\crv$, which is an integer given by the intersection
\be
\degr{\crv} := -\candl \cdot \crvl \,,
\ee
where $\candl$ is the pull-back of the canonical divisor of the F-theory base $\fthb$ to the ambient space, $\asl$, of the resolved D7-brane, as described in detail in Appendix~\ref{app:tadpole}.

We note that Equation~\eqref{eq:tadp_contr_gen} reduces for  $g_\crv=0$ to the result in Equation~\eqref{eq:tadp_cont_simpl_ex} for the simplest example (for which the anti-canonical divisor of the base $\fthb$ is equal to $4H$ where $H$ is the hyperplane class of $\mbb{P}^3$, so that the expressions are related by $\degr{} = 4d$) and also reduces to the Equation~\eqref{eq:tadp_cont_tor} that we derive for toric weak Fano bases in Appendix~\ref{app:toric}.

\bigskip

When the genus of the flux curve $\crv$ is zero and the F-theory base is simply a $\mbb{P}^3$, the embedding of the curve can be specified by polynomial maps, making the computation of the number of stabilized moduli straightforward. However, for more generic configurations the embedding map is not polynomial and is not given explicitly. In Appendix~\ref{app:mod_count} we regale the interested reader with all the details of this computation. The conclusion is that there is the following upper bound for the number of stabilized moduli,
\be
\stbmod{\crv}\leq  8\degr{\crv} + 1 \,.
\label{eq:mod_stab_gen}
\ee
We would like to note that obtaining this very general result was made possible by a rather unexpected stroke of luck: there are a number of terms in the computation which appear particularly difficult to evaluate for generic base spaces. However, we were able to show that, if present, these terms would only {\it decrease} the number of stabilized moduli. This allowed us to derive the bound above without explicitly evaluating these terms\footnote{If any of these terms had the opposite sign, it would have been impossible to derive a general result for all compactifications with a weak Fano base.}.

We can see that Equation~\eqref{eq:mod_stab_gen} reduces to the simpler expressions obtained for less generic base spaces:  Equation~\eqref{eq:mod_stab_p3} for ${\mathbb P}^3$ and Equation~\eqref{eq:mod_stab_tor} for generic toric bases. 

It is notable that, unlike the expression for the D3 charge contribution $\flxcont{\crv}$, the bound on the number of stabilized moduli  $\stbmod{\crv}$, does not have a dependence on the genus $g_\crv$.

\subsubsection*{The window of worldvolume fluxes that are not excluded}
\label{sec:window}

Having obtained the bound on the D3 charge sourced by the D7 worldvolume fluxes and the bound on the total number of stabilized D7 moduli, we can now compare these to the values of the total D3 tadpole and the total number of D7-brane moduli. The conditions that this flux is smaller than the tadpole bound and completely stabilizes the D7-brane moduli are
\be
\begin{aligned}
\mathrm{Tadpole~cancelation:}~&~ &&(1-g_\crv) + \tfrac72 \degr{\crv} && \leq && Q_\mathrm{neg.} && = && 12 + 15C_1^3 \,, \\
\mathrm{Stabilization:}~&~ && 8\degr{\crv} + 1 && \geq && n_{\mathrm{D}7} && = && 16 + 58C_1^3 \,.
\end{aligned}
\label{eq:constraints}
\ee
where we have recalled the computations of the D3 charge tadpole $Q_\mathrm{neg.}$ and of the number of D7-brane moduli $n_{\mathrm{D}7}$ from Equations~\eqref{eq:tadp_gen} and \eqref{eq:nD7}, and where we have used the shorthand $C_1^3 \equiv \int_{\fthb}c_1(\fthb)^3$.

\begin{figure}[h]
\includegraphics[scale=.7]{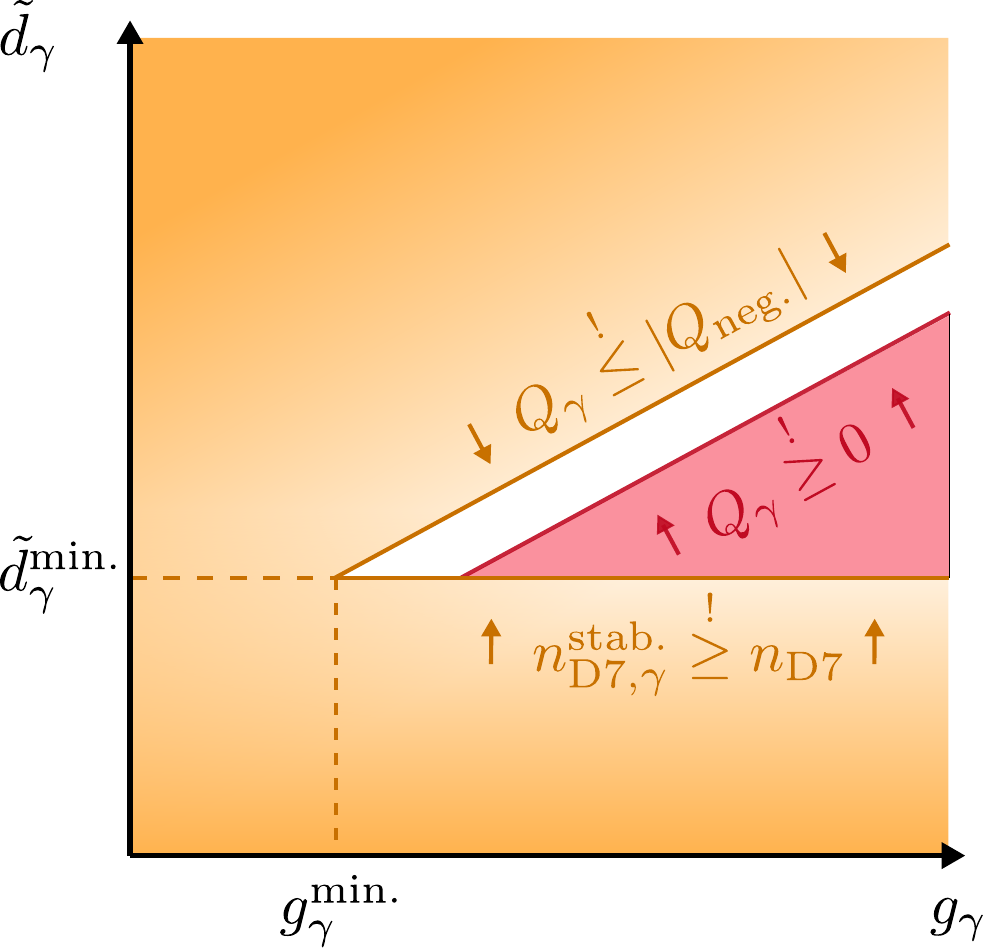}%
\caption{Orange: The degrees $\degr{\crv}$ and genera $g_\crv$ of flux curves $\crv$ that our calculations rule out as leaving unfixed  D7 moduli or as incompatible with tadpole cancelation. Red: Fluxes that our calculation does not rule out but which give a negative tadpole contribution and are hence unphysical. White: Fluxes that we cannot rule out using our calculation.}
\label{fig:constraint_plot}%
\end{figure}

We plot these two constraints in orange in \fref{fig:constraint_plot}, in which the white and red regions correspond to the curves that are not ruled out by our calculations\footnote{One may wonder if there exist purely mathematical relations between the canonical degree, $C \cdot \cand{S}$, and the genus, $g_C$, of a curve on a surface $S$. Such a direct relation exists for $S = \mbb{P}^2$, but for more general surfaces there is only an inequality. Specifically, using $g_C = \frac{1}{2}(C^2 + C \cdot \cand{S} +2)$ one can see that the genus is maximized when $C \propto \cand{S}$, so that in particular $C = \frac{C \cdot \cand{S}}{\cand{S}^2}$ and hence $g_C^\mathrm{max} = \frac{1}{2}\left(\frac{(C \cdot \cand{S})^2}{\cand{S}^2}+C \cdot \cand{S} + 2 \right)$. However this inequality lies below the white region in \fref{fig:constraint_plot}, and so does not provide an additional constraint.}. The gradients of the two constraint lines are fixed, and are independent of the choice of F-theory base space, while the tip of the triangular region that we do not rule out, whose coordinates are $\big( g_\crv^\mathrm{min.} \,,\, \degr{\crv}^\mathrm{min.} \big) $, moves up and to the right as one increases the value of $C_1^3$ (which increases the number of moduli):
\be
\big( g_\crv^\mathrm{min.} \,,\, \degr{\crv}^\mathrm{min.} \big) = \Big( \tfrac{1}{16}(166C_1^3-71) \,,\, \tfrac{1}{8}(58C_1^3+15) \Big) \,.
\label{eq:min_g_and_d}
\ee

Note that our calculations can only be used to rule out flux vacua, but this does not imply that the flux vacua that we do not rule out are physical. Indeed, the triangular region allowed by our two constraints includes flux configurations that would give a negative contribution to the flux tadpole, and we have shown these configurations in red. These have not been explicitly forbidden in our analysis because we have only imposed a necessary but not sufficient condition for anti-self-duality of the flux, as we discussed below Equation~\eqref{eq:flux_form}. It is clear that such configurations are not allowed: Anti-self-dual fluxes must give a positive contribution to the D3 tadpole everywhere in the geometry, which cannot integrate to a negative value. Hence, configurations with a total tadpole contribution that is negative are incompatible with a metric that makes the fluxes anti-self-dual. It is quite likely that enforcing the positivity of the tadpole contribution of the fluxes {\em locally} on the D7-brane,
and also imposing the primitivity condition at the level of the actual forms and not only at the level of cohomology classes, will lead to even stronger bounds. We leave this exploration for later work.

\medskip

For a given fixed F-theory base space, the above bounds give a no-go theorem for D7 moduli stabilization across the vast swathe of worldvolume fluxes dual to curves outside of the white region in \fref{fig:constraint_plot}.

What values can the vertex $( g_\crv^\mathrm{min.} \,,\, \degr{\crv}^\mathrm{min.})$ of the non-excluded region take? Since the F-theory base, $\fthb$, is assumed to be weak Fano, $C_1^3 \equiv \int_{\fthb}c_1(\fthb)^3$ is a positive integer. Hence, from the expressions in Equation~\eqref{eq:min_g_and_d}, we see that $g_\crv^\mathrm{min.}$ and $\degr{\crv}^\mathrm{min.}$ are always greater than zero. Notably, this observation extends to all possible weak Fano base spaces the no-go theorem for genus-zero flux curves derived for a $\mbb{P}^3$ base space in Ref.~\cite{Collinucci:2008pf} and reviewed in Section~\ref{sec:nogo_review}.

Further, the value of $C_1^3$ is always even for a weak Fano three-fold, as can be seen for example from the expression in Equation~\eqref{eq:acbund_ind}. Hence the lowest possible value is $C_1^3 = 2$. This value is in fact realised, for example by the hypersurface of degree 6 in the weighted projective space $\mbb{P}^4_{1,1,1,1,3}$. For such a three-fold base, with $C_1^3 = 2$, one has
\be
\big( g_\crv^\mathrm{min.} \,,\, \degr{\crv}^\mathrm{min.} \big) \geq ( 17 , 17 ) \,.
\ee
Hence, for any weak Fano base space, it is not possible to stabilize the D7 brane moduli with flux dual to a curve with genus less than $17$.

Finally, for the particular choice $\fthb = \mbb{P}^3$ treated in Ref.~\cite{Collinucci:2008pf}, one has $C_1^3 = 64$, which gives
\be
\big( g_\crv^\mathrm{min.} \,,\, \degr{\crv}^\mathrm{min.} \big) \geq \big(660 \,,\, 466) \,.
\ee
Thus, our calculation rules out stabilization within the tadpole across a much wider range of fluxes than the genus-zero no-go theorem of Ref.~\cite{Collinucci:2008pf}.

\subsubsection*{Fluxes dual to curves of high degree and high genus}

It is interesting to understand the physics of the flux curves in the white sliver that is not ruled out by our calculation. For compactifications with a small value of $C_1^3 \equiv \int_{\fthb}c_1(\fthb)^3$, some of the fluxes that are not excluded have small degree and genus, and hence do not appear pathological. Hence, our calculation leaves open the possibility that small numbers of D7-brane moduli can be stabilized within the tadpole bounds. 

However, in the spirit of the Tadpole Conjecture, we are interested in the opposite limit, when the number of D7 moduli becomes very large. In that limit, our arguments do not rule out certain families of curves of high degree and high genus that, if physical, could stabilize a growing number of moduli within the tadpole bounds, and hence violate the Tadpole Conjecture. However, while the calculations of the present paper do not exclude such configurations, there is reason to believe that these configurations may be problematic in their own right. 

To see this, recall from the calculation above that as the number of D7 moduli grows, the flux required to stabilize the D7-brane is dual to a curve of ever-increasing genus. But as the genus increases, the variation of the profile of the flux becomes more and more non-trivial. Hence, one can expect that at some point the scale of variation of the flux will dip below the string scale, at which point the supergravity description breaks down. Said differently, to accommodate fluxes with such a rapidly varying profile, the volume of the compactification would need to be very large, eventually taking on phenomenologically excluded values.

Hence, in the limit of many D7 moduli, one can expect  phenomenological problems to arise with the flux required to stabilize the D7-brane. We plan to return to this argument in future work, to flesh out and quantify the details.

\section{Conclusions}
\label{Conclusions}

We have examined the stabilization of D7-brane moduli by turning on worldvolume fluxes, dual to a curve $\crv$. We have derived a lower bound on the D3 charge induced by these fluxes and an upper bound on the number of D7 moduli they stabilize, for F-theory compactifications with a weak Fano base. For fluxes dual to genus-zero curves, which is the only case previously considered in the literature in this context, our bounds imply that for a large number of moduli, the ratio between the charge sourced by the fluxes and the number of moduli these fluxes stabilize is 
\be
\alpha\equiv \frac{Q_{\crv}}{\stbmod{\crv}} \ge \frac{7}{16}=0.4375. \nonumber \ 
\ee
This result applies both to compactifications over any of the $O(10^{15})$ toric weak Fano bases \cite{Halverson_2017} and also to more general compactifications whose weak Fano base is not necessarily toric\footnote{The number of such bases can be expected to be much larger, but to our knowledge has not been estimated.}, and spectacularly confirms the Tadpole Conjecture \cite{Bena:2020xrh}, both by the linear growth of the tadpole sourced by the fluxes with the number of moduli they stabilize,\footnote{For more recent work on the linear relation between the tadpole sourced by the fluxes that stabilize moduli and the number of moduli see Refs.~\cite{Plauschinn, Severin} in the context of a stabilization proposal put forth in Ref.~\cite{Marchesano}.}  and also by the fact that the proportionality coefficient, $\alpha$, is larger than $1/3$. Since this is larger than 0.259, the ratio allowed by the tadpole cancelation condition (see \eqref{tadpolenD7}), D7 moduli cannot be stabilized by fluxes dual to a curve of genus zero, within the tadpole bound.  

We have also considered flux curves whose genera are large, and we have found that increasing the genus while keeping the degree fixed weakens our bounds on the minimum tadpole contribution of the fluxes, as one can see in Equation~\eqref{eq:constraints}. Thus, if one considers the general degree-genus plane, depicted in Figure~\ref{fig:constraint_plot}, our bounds do not exclude the possibility that moduli stabilization within the tadpole bound may be possible in a certain region of high-degree high-genus curves. Nevertheless, this does not imply that all the flux curves we do not rule out are physical; in fact, it is easy to immediately see that a large part of these fluxes (depicted in red in Figure~\ref{fig:constraint_plot}) give rise to negative D3-brane charge, and hence are unphysical.
The reason why our calculation does not rule them out is likely because we  only imposed the primitivity of the fluxes at the level of cohomology, so many of the flux configurations that escape our bounds are in fact non-supersymmetric.

The inability to exclude fluxes that give a negative tadpole contribution is an illustration of the fact that our calculation is only concerned with topology and fluxes, but does not address the harder question whether there exists a metric sourced by the fluxes that is physical. Ruling out fluxes whose overall tadpole contribution is negative is a clear first step of including the metric in our considerations: If there existed a metric for which the fluxes are anti-self-dual (as required by supersymmetry) their tadpole contribution would be positive, and hence could not integrate to a negative number; hence no such metric exists and the flux curves in the red region of Figure~\ref{fig:constraint_plot}, which give a negative tadpole contribution, must be excluded. Furthermore, it is not clear whether this similar considerations would not rule out the rest of the flux curves in the white sliver of Figure~\ref{fig:constraint_plot}.

Hence, despite the suggestiveness of the coherence of the no-go theorem for small genus with the Tadpole Conjecture, our calculations do not rule out the possibility of stabilizing large numbers of D7 moduli within the tadpole bound, nor do they give us a proof of the Tadpole Conjecture everywhere in the degree-genus plane. As we explained above, such a proof might require more detailed considerations of the metric sourced by the fluxes, and we believe that, ultimately, it may be possible to argue that the sliver of configurations that our calculation does not exclude, is still problematic. 

We have also argued that for a very large number of D7 moduli, the flux required to stabilize the D7-brane can be expected to have a scale of variation eventually dipping below the string scale, unless the volume of the compactification manifold is larger than values compatible with the real world. Hence, in the  limit of a very large number of moduli, we believe it is impossible to stabilize the D7-brane moduli while keeping the curvature sub-stringy -- a result that agrees with the Tadpole Conjecture, albeit not in the string-scale-independent way it was formulated in Ref.~\cite{Bena:2020xrh}.

\medskip

\vspace{0.6cm}
\noindent {\bf Acknowledgements:} We would like to thank Jim Halverson and Severin L\"ust for useful discussions. This work was supported in part by the ANR grant Black-dS-String ANR-16-CE31-0004-01, by the ERC Grants 772408 ``Stringlandscape'' and 787320 ``QBH Structure'', by the John Templeton Foundation grant 61149, and by the NSF grant PHY-2014086.

\appendix

\section{The geometry and the D7-branes}
\label{app:geom}

Here we compute some properties of the background geometry and D7-brane configuration of Section~\ref{sec:setup_geom} which we stated there without proof. We note that some of the quantities below have also been computed from a slightly different viewpoint in Section~5 of Ref.~\cite{Collinucci:2008pf}.

As discussed in Section~\ref{sec:setup_geom}, we restrict to compactifications where the F-theory four-fold $Z_4$ is smooth and of strict SU(4) holonomy. Moreover, we take the four-fold to be given by a Weierstrass model with a single section over a base $\fthb$ which is weak Fano\footnote{A manifold is weak Fano if the anti-canonical bundle $\canb{\fthb}^*$ is nef and big. In terms of the anti-canonical divisor $-\cand{\fthb}$, this means that $(-\cand{\fthb}) \cdot C \geq 0$ for any curve $C$ and $(-\cand{\fthb})^n > 0$, with $n$ the dimension of the manifold.}. The condition that the base be weak Fano is closely related to smoothness of the four-fold, but there are subtleties in both directions, as discussed in Section~3.5.1 of Ref.~\cite{Halverson:2015jua}.

We first compute the D3 tadpole. When $Z_4$ is smooth, the Euler characteristic of the four-fold $Z_4$ is simply related to quantitites on the base via \cite{Sethi:1996es}
\be
\tfrac{1}{24}\ec(Z_4) = \int_{\fthb}\big(15c_1(\fthb)^3+\tfrac{1}{2}c_1(\fthb)c_2(\fthb)\big) \,.
\ee
When the holonomy of $Z_4$ is $\mathrm{SU}(4)$, the Hodge numbers, $h^{i,0}(Z_4)$, vanish for $i=1,2,3$. But if the base had non-zero Hodge numbers $h^{i,0}(\fthb)$, these forms would pull-back to $Z_4$. Hence, the same vanishings must hold on the base $\fthb$. Since $h^{i,0}(\fthb) \equiv h^i(\fthb,\mc{O}_{\fthb})$, this implies that $\ind\big(\mc{O}_{\fthb}\big) = 1$. But on any three-fold, $\ind\big(\mc{O}_{\fthb}\big) = \tfrac{1}{24}\int_{\fthb} c_1(\fthb)c_2(\fthb)$. Hence, $\int_{\fthb} c_1(\fthb)c_2(\fthb) = 24$. This simplifies the expression of $\ec(Z_4)$ and of the corresponding negative contribution to the D3 tadpole: 
\be
Q_\mathrm{neg} = - \frac{\ec(Z_4)}{24} = - \left(12 + 15\int_{\fthb}c_1(\fthb)^3\right) \,.
\ee

Next we compute the number of deformation moduli of the D7-brane. We recall from Equation~\eqref{eq:num_d7_mod_gen} that for a general base space these are counted by
\be
n_{\mathrm{D}7}=h^0\big(\fthb,\,(\canb{\fthb}^*)^{ 4}\big) + h^0\big(\fthb,\,(\canb{\fthb}^*)^{6}\big) - h^0\big(\fthb,\,(\canb{\fthb}^*)^{2}\big) - 1 \,.
\ee
Since $\canb{\fthb}^*$ is nef and big, its higher cohomologies (and those of its positive powers, $(\canb{\fthb}^*)^{\otimes \alpha}$ for $\alpha \geq 0$) vanish by the Kawamata-Viehweg vanishing theorem (see for example Chapter~9.1.C of Ref.~\cite{lazarsfeld2004positivity2}). Hence, the above zeroth cohomologies can be computed with the index, $h^0\big((\canb{\fthb}^*)^{\otimes \alpha}\big) = \ind\big( (\canb{\fthb}^*)^{\otimes \alpha} \big)$. Additionally, the latter is easily computed using the Atiyah-Singer index theorem,
\be
\begin{aligned}
\ind\big( (\canb{\fthb}^*)^{\otimes \alpha} \big)
& = \int_{\fthb} \mathrm{ch}\big( (\canb{\fthb}^*)^{\otimes \alpha} \big) \w \mathrm{Td}(\tanb{\fthb}) \\
&= \int_{\fthb} (1+\alpha c_1 + \tfrac{1}{2}\alpha^2 c_1^2+ \tfrac{1}{6}\alpha^3C_1^3) \w \big(1+\tfrac{1}{2}c_1+\tfrac{1}{12}(c_1^2+c_2)+\tfrac{1}{24}c_1c_2\big) \\
&=  \tfrac{\alpha}{12}(1+\alpha)(1+2\alpha)\int_{\fthb}c_1(\fthb)^3+\tfrac{1}{24}(1+2\alpha)\int_{\fthb}c_1(\fthb)c_2(\fthb) \\
&= (1+2\alpha) + \tfrac{\alpha}{12}(1+\alpha)(1+2\alpha)\int_{\fthb}c_1(\fthb)^3 \,,
\end{aligned}
\label{eq:acbund_ind}
\ee
where in the first term of the final line we have again used the fact that $\int_{\fthb} c_1(\fthb)c_2(\fthb) = 24$. Plugging in  the values $\alpha = 4,6,2$, we hence find that
\be
n_{\mathrm{D}7} = 16 + 58\int_{\fthb}c_1(\fthb)^3 \,.
\ee
We also note that, since the base $\fthb$ is weak Fano, the anti-canonical bundle is big, so $\int_{\fthb} c_1(\fthb)^3$ is a positive integer. Moreover this integer is an even integer, as is clear from the relation in Equation~\eqref{eq:acbund_ind} for e.g.\ $\alpha=1$.

\section{The D3 charge sourced by the D7 fluxes}
\label{app:tadpole}

In the main text we have used the expression of the contribution of the D7 worldvolume flux, $F$ to the D3-brane tadpole
\be
Q_F = - \frac{1}{4}\int_\db F \w F \,.
\ee

Evaluating this integral is complicated by the fact that the surface $\db$ wrapped by the D7-brane is singular, generically having double-point intersections \cite{Collinucci:2008pf,Braun_2008}. To proceed, one can blow-up along the singularities to give a smooth resolved surface, $\dbl$, lifting also the flux to a (1,1)-form $\overline{F}$ on $\dbl$, and there compute the integral $Q_F = - \frac{1}{4}\int_{\dbl} \overline{F} \w \overline{F}$.

We first recall the construction of the resolved D7-brane surface, $\dbl$, following Ref.~\cite{Collinucci:2008pf}, and then turn to the computation of the integral $\int_{\dbl} \overline{F} \w \overline{F}$ for fluxes of the form considered in the main text: $F \sim \crv - \crv'$ where $\crv$ is a holomorphic curve that the D7-brane has been tuned to contain, and $\crv'$ its orbifold image.

\subsubsection*{The resolved D7-brane}

At the intersection between the D7-brane and the O7-plane, whose geometries have been discussed in Section~\ref{sec:setup_geom}, the D7-brane intersects itself, so that the wrapped surface, $\db$, is singular. In the notation of Section~\ref{sec:setup_geom}, this intersection occurs where $\eta = \xi = 0$, which describes a curve inside the double-cover CY.

By blowing up along this curve, the two branches of the D7-brane are separated, removing the singularities and giving rise to a smooth surface, $\dbl$. Practically, this resolution may be performed by blowing up the ambient space, $\as$, of the double-cover CY described by Equation~\eqref{eq:cy_eqn} to give a new ambient space, $\asl$, and taking the proper transform $\dbl$ of the D7-brane surface.

In Ref.~\cite{Collinucci:2008pf} the authors construct explicitly the resolved D7-brane surface $\dbl$ for the simple F-theory compactification discussed in Section~\ref{sec:nogo_review}, and also determine a number of properties of general compactifications. In particular, it is straightforward to show that the (object that restricts from $\asl$ to $\dbl$ to give the) first Chern class of $\dbl$ is given by
\be
c\big(\asl,\tanb{\dbl}\big) = 1 + \big([O]-[D]\big) + \ldots \,,
\ee
where $[O]$ and $[D]$ are respectively the classes of the O7-plane and the D7-brane, lifted to the new ambient space $\asl$. Recalling from Section~\ref{sec:setup_geom} that the O7-plane and D7-brane respectively correspond to sections of $(\canb{\fthb}^*)^{\otimes 1}$ and $(\canb{\fthb}^*)^{\otimes 8}$, we see that
\be
c\big(\asl,\tanb{\dbl}\big) = 1  - 7[-\candl] + \ldots \,,
\ee
where we have written $\candl$ for the pull-back to $\asl$ of the canonical divisor of $\fthb$ through the composition of the projection map from $\as$ to $\fthb$ and the blow-down map from $\asl$ to $\as$.

\subsubsection*{The tadpole contribution}

We would like to calculate the D3-brane charge sourced by worldvolume fluxes of the form $F \sim \crv - \crv'$, with $\crv$ an irreducible curve inside the tuned D7-brane, $\db$, and $\crv'$ its orbifold image. These naturally lift in the resolution of $\db$ to curves $\crvl$ and $\crvl'$ on the smooth surface $\dbl$. Moreover, the D3 tadpole contribution of the worldvolume flux $F$ can be computed as
\be
\flxcont{\crv} = - \frac{1}{4}(\crvl - \crvl') \cdot (\crvl - \crvl') = \frac{1}{2}\big(-\crvl^2 + \crvl \cdot \crvl'\big) \,.
\label{Q-bound}
\ee
Since $\crv$ and $\crv'$ are curves inside the CY three-fold $\cyu$ related by the orbifold action, we expect intersections only at the orbifold fixed locus, which are generically separated by the resolution of the D7-brane, and we hence expect that $\crvl \cdot \crvl' = 0$. If this intersection happens to be non-zero, then, since it is an intersection between distinct irreducible curves, it must be positive. Hence Equation~\eqref{Q-bound} certainly gives a lower bound: $\flxcont{\crv} \geq -\crvl^2$.

Noting that this first term can be rewritten using the relation
\be
-\crvl^2 = \chi(\crvl)- c_1(\dbl,\tanb{\dbl}) \cdot \crvl \,,
\ee
using the above expression for $c_1(\asl,\tanb{\dbl})$ and writing $g_\crv$ for the genus of $\crv$, we see that 
\be
\flxcont{\crv} \geq -\frac{1}{2}\,\crvl^2 = (1-g_\crv) + \frac{7}{2} (-\candl) \cdot \crvl  \equiv (1-g_\crv) + \frac{7}{2} \degr{\crv} \,,
\label{eq:general_q}
\ee
where the intersection in the second term is taken on $\asl$, and where we have defined the canonical degree associated to the curve: $\degr{\crv} \equiv (-\candl) \cdot \crvl$ .

\section{Counting the stabilized moduli}
\label{app:mod_count}

Here we consider the problem of counting the number of constraints on the D7-brane moduli space that result from demanding that the D7-brane contain a particular irreducible curve.

\bigskip

We recall from Section~\ref{sec:setup_geom} that the D7-brane locus is determined by the intersection between the CY hypersurface $\xi^2=h(x)$ and the D7-brane equation $\eta(x)^2 = h(x)\xi(x)$. The D7-brane equation is a section of\footnote{Here and in the following we abuse notation by writing $\cand{\fthb}$ for the pull-back to $\as$ of the canonical divisor on $\fthb$ by the projection map from $\as$ to $\fthb$.} the bundle $\mc{O}_\as(-8\cand{\fthb})$, but its form is not the most general one, so the possible configurations of the D7-brane correspond to a particular subspace of the space of sections $\Gamma\big(\mc{O}_\as(-8\cand{\fthb})\big)$. In demanding that the D7-brane contain some curve $\crv$, we are imposing a second constraint which will cut this subspace down further. In general, there may be a non-trivial interplay between these two constraints on the space of sections. However, such an interplay can only result in a smaller number of coefficients being constrained. Hence, if we assume the best-case scenario for stabilization, we can consider the two constraints independently, and simply sum the number of coefficients being constrained by each. And, since the number of coefficients constrained by demanding a form $\eta(x)^2 = h(x)\xi(x)$ has already been taken into account in the original counting of D7-brane deformations in Equation~\eqref{eq:num_d7_mod_gen}, it remains only to determine the dimension of the subspace of elements in $\Gamma\big(\mc{O}_\as(-8\cand{\fthb})\big)$ which contain the curve $\crv$. Hence, an upper bound on the number of constraints $\stbmod{\crv}$ on the D7-brane from demanding the inclusion of the curve $\crv$ is given by
\be
\stbmod{\crv} \leq 
\mathrm{dim}\,\left(\Gamma\big(\mc{O}_\as(-8\cand{\fthb})\big)\right) - \mathrm{dim}\,\big\{s \in \Gamma\big(\mc{O}_\as(-8\cand{\fthb})\big) \, | \,  s \stackrel{!}{\supset} \crv \big\} \,.
\ee

It is in general difficult to compute a quantity like the second term, which is the dimension of the subspace of sections which contain a particular locus in the manifold. One way to proceed is to note that, if we blow up along the specified sublocus $\crv$ to give an exceptional divisor, $E$, on the blown-up manifold $\asl$, the sections of the bundle $\mc{O}_{\asl}(D-E)$ are in one-to-one correspondence with the sections of $\mc{O}_\as(D)$ which vanish along $\crv$. This follows from the fact that $\mc{O}(D)$ can be seen as the sheaf of functions whose divisor is greater or equal to $D$.
Writing $\candl$ for the pull-back to $\asl$ of the canonical divisor of $\fthb$ through the composition of the projection map from $\as$ to $\fthb$ and the blow-down map from $\asl$ to $\as$, and writing $E_\crv$ for the exceptional divisor that blows-down to $\crv$, we can hence rewrite the above upper bound as
\be
\stbmod{\crv} \leq h^0\left(\mc{O}_{\asl}\big(-8\candl\big)\right) - h^0\left(\mc{O}_{\asl}\big(-8\candl-E_\crv\big)\right) \,,
\ee
where we have also chosen to replace the first term using the equality between $h^0\left(\mc{O}_{\as}\big(-8\cand{\fthb}\big)\right)$ and $h^0\left(\mc{O}_{\asl}\big(-8\candl\big)\right)$.

The expression for the upper bound is now a difference of cohomologies on the same variety $\asl$. To compute this, we first note that on a variety $Y$, given a divisor $D$, there is the exact sequence of sheaves
\be
0 \to \mc{O}_Y(-D) \to \mc{O}_Y \to  \mc{O}_Y|_D \to 0 \,.
\ee
To make use of this sequence we set $Y = \asl$ and take $D = E_\crv$, so that we have
\be
0 \to \mc{O}_{\asl}(-E_\crv) \to \mc{O}_{\asl} \to  \mc{O}_{\asl}|_{E_\crv} \to 0 \,,
\ee
and then tensor the sequence with $\mc{O}_{\asl}\big(-8\candl\big)$, to give
\be
0 \to \mc{O}_{\asl}\big(-8\candl-E_\crv\big) \to \mc{O}_{\asl}\big(-8\candl\big) \to \mc{O}_{\asl}\big(-8\candl\big)\big|_{E_\crv} \to 0 \,.
\ee
To make the connection between this sequence of sheaves and the required cohomologies, we first take the associated long exact sequence in cohomology, which reads
\be
\begin{aligned}
0 &\to H^0\Big(\mc{O}_{\asl}\big(-8\candl-E_\crv\big)\Big)
\to H^0\Big(\mc{O}_{\asl}\big(-8\candl\big)\Big)
\to H^0\Big(\mc{O}_{\asl}\big(-8\candl\big)\big|_{E_\crv}\Big)
\\
&\to H^1\Big(\mc{O}_{\asl}\big(-8\candl-E_\crv\big)\Big)
\to H^1\Big(\mc{O}_{\asl}\big(-8\candl\big)\Big)
\to H^1\Big(\mc{O}_{\asl}\big(-8\candl\big)\big|_{E_\crv}\Big)
\\
&\to ~~\ldots 
\end{aligned}
\ee
Recalling the discussion in Appendix~\ref{app:geom}, we note that the higher cohomologies of $\canb{\fthb}^*$ (and its positive tensor powers) vanish, and hence the long exact sequence terminates after only four terms. Hence, the alternating sum of the dimensions of these four terms must vanish, or rearranging,
\be
\begin{aligned}
&~ &&h^0\left(\mc{O}_{\asl}\big(-8\candl\big)\right) - h^0\left(\mc{O}_{\asl}\big(-8\candl-E_\crv\big) \right) \\
&= && h^0\left(\mc{O}_{\asl}\big(-8\candl\big)\big|_{E_\crv}\right) - h^1\left(\mc{O}_{\asl}\big(-8\candl-E_\crv\big) \right) \,.
\end{aligned}
\ee
The first line is precisely the quantity we want to compute. In the second line, the second term will only reduce the number of stabilized moduli, and hence we have the inequality
\be
\stbmod{\crv} \leq h^0\left(\mc{O}_{\asl}\big(-8\candl\big)\big|_{E_\crv}\right) \,.
\ee

It remains to compute this final quantity, which is the zeroth cohomology of a line bundle on the complex surface $E_\crv$. To compute this, we note that $E_\crv$ is a fiber-bundle of $\mbb{P}^1$ over the base curve $\crv$. Notably, the line bundle on $E_\crv$ is clearly a pull-back to $E_\crv$ of a line bundle on the base $\crv$ with degree $-8\candl \cdot \crvl$. Furthermore, the zeroth cohomology of the pull-back bundle on $E_\crv$ is simply given by the zeroth cohomology of the bundle on $\crv$. Hence, it remains only to compute the zeroth cohomology of a line bundle on the curve $\crv$.

When the genus $g_\crv$ of the curve $\crv$ is non-zero, the zeroth cohomology of a line bundle on $\crv$ is not determined uniquely by the degree. It is however bounded from above by the degree plus one (see for example Theorem~9.6\,(i) of Ref.~\cite{BogomolovPetrov2002}). Hence
\be
h^0\left(\mc{O}_{\asl}\big(-8\candl\big)\big|_{E_\crv}\right) \leq -8\candl \cdot \crvl + 1 \,,
\ee
This gives finally an upper bound on the number of constraints on the D7-brane that result from requiring the inclusion of a particular complex curve $\crv$, 
\be
\stbmod{\crv} \leq - 8\candl \cdot \crvl + 1  \equiv 8\degr{\crv} + 1 \,,
\ee
where we have defined the canonical degree $\degr{\crv} \equiv -\candl \cdot \crvl$.

\section{Toric base space and genus-zero flux}

In Section~\ref{sec:nogo_review} we reviewed the no-go theorem of Ref.~\cite{Collinucci:2008pf}, which applies only when the F-theory base is a $\mbb{P}^3$ and the stabilizing worldvolume flux is dual to a curve of genus zero. In this appendix we extend this no-go theorem to the situation where the F-theory base is any toric weak Fano three-fold, keeping the flux dual to a curve of genus zero. In Section~\ref{sec:gen} we have treated a much more general scenario, loosening both the toric restriction and the genus-zero restriction. However while the geometries we consider in this Appendix are less generic, they are amenable to computations which are more explicit and illustrative. 

\subsubsection*{Toric geometry}
\label{app:toric}

When the F-theory base, $\fthb$, is a toric variety, it is specified by a weight system, which we can write in general as
\be
\fthb \sim 
\begin{array}{c c c}
x_0 & \ldots & x_n \\
\hline
q_1^0 & \cdots & q_1^n \\
\vdots & \ddots & \vdots \\
q_m^0 & \cdots & q_m^n 
\end{array} ~,
\label{eq:fthb_ws}
\ee
as well as a Stanley-Reisner ideal, which we do not write explicitly. We note that the anti-canonical bundle $\canb{\fthb}^*$ of $\fthb$ is the line bundle with charge $\sum_{j=1}^n q_i^j$ under the $i$th row of the charge matrix. For example, if $\fthb = \mbb{P}^3$, then the weight system is
\be
\fthb \sim 
\begin{array}{c c c c}
x_0 & x_1 & x_2 & x_3 \\
\hline
1 & 1 & 1 & 1
\end{array} ~,
\ee
and $\canb{\fthb}^*$ is the line bundle with charge $4$ (i.e.\ whose sections correspond to degree 4 polynomials) under the projective scaling $x_i \to \lambda x_i$ where $\lambda \in \mbb{C}^*$.

The double-cover of the orientifold in the Type IIB perspective is then described by the hypersurface in Equation~\eqref{eq:cy_eqn} inside a toric ambient space $\as$ with weight system
\be
\as \sim
\begin{array}{c c c c}
x_0 & \ldots & x_n & \xi \\
\hline
q_1^0 & \cdots & q_1^n & \sum_j q_1^j\\
\vdots & \ddots & \vdots & \vdots \\
q_m^0 & \cdots & q_m^n & \sum_j q_m^j
\end{array} ~.
\ee
in which the charges for $\xi$ have been chosen to make Equation~\eqref{eq:cy_eqn} consistent. We note that $\as$ is singular, but also that the singularities generically miss the hypersurface, so that for many computations it is not necessary to resolve the singularities of $\as$.

We define a basis of divisor classes on $\as$ by associating a class, $H_i$, to the unit charge in each row of the weight system. For example, when $\fthb = \mbb{P}^3$
\be
\as _{\mbb{P}^3} \sim
\begin{array}{c c c c c c}
x_0 & x_1& x_2 & x_3 & \xi \\
\hline
1 & 1 & 1 & 1 & 4
\end{array} ~,
\ee
and the basis of divisors is given by a single element, $H \sim [x_0] \sim \ldots \sim [x_3]$, which is the hyperplane class of the weighted projective space $\as_{\mbb{P}^3} = \mbb{P}^4_{1,1,1,1,4}$.

\subsubsection*{Fluxes and moduli stabilization}

We stabilize the D7-brane moduli following the procedure outlined in Section~\ref{sec:setup_stab}. Here we assume, as in the simple example treated in Section~\ref{sec:nogo_review}, that the flux curve $\crv$ has genus zero. This has the distinct advantage that both the embedding space $\as$ and the curve itself are covered by homogeneous coordinates, which we can use to parameterize the embedding, to hence easily count how many D7-brane moduli are stabilized.

\medskip

Since the flux curve $\crv$ is a $\mbb{P}^1$, we may parameterize it with homogeneous coordinates $[u:v]$, and embed the curve $\crv$ into the ambient space $\as$ of the upstairs double-cover $\cyu$ with polynomial maps:
\be
x_i = X_i(u,v) \,, ~ \xi = \Xi(u,v) \,,
\ee
where $X_i(u,v)$ and $\Xi(u,v)$ are homogeneous polynomials in $[u:v]$. In order for this map to be consistent (not one-to-many), the degrees of these polynomials must be consistent with the allowed projective scalings of the coordinates $x_i$ and $\xi$, as specified by the toric weight system of $\as$. One can check that this requires that
\be
\mathrm{deg}(X_i) = \sum_j q_j^i d_j \,, \quad \mathrm{deg}(\Xi) = \sum_{i,j} q_j^i d_j \,,
\ee
where $d_i$ are arbitrary integers we associate to each row of the weight system of $\as$, and the $q_j^i$ are the entries in this weight system. Importantly, the choice of the integers $d_i$ corresponds to the curve class of $\crv$ having the intersection properties
\be
\crv \cdot H_i = d_i \,.
\label{eq:crv_ints}
\ee
The combination $\sum_{i,j} q_j^i d_j$ will reoccur frequently below, so it will be useful to define
\be
\degr{\crv} \equiv \sum_{i,j} q_j^i d_j \,.
\ee

The embedding map is specified by the coefficients in the polynomials $X_i$ and $\Xi$. However, some maps are equivalent. First, there are the $n-2$ projective scaling identifications on the coordinates of $\as$, as specified by the weight system in Equation~\eqref{eq:fthb_ws}, which identify naively distinct embedding maps, reducing the true number of parameters by $n-2$. Second, the $\mbb{P}^1$ admits a family of $\mathrm{GL}_2$ reparametrizations, which is a 4-dimensional group of transformations. However, one of the latter corresponds to an overall scaling identification, which overlaps with the scaling identifications on the toric target space. Hence, the true number of parameters of the embedding map is
\be
\sum_{i=0}^n \big(\mathrm{deg}(X_i)+1\big) + \big(\mathrm{deg}(\Xi)+1\big) - (n-2) - 3 = 2\degr{\crv}  + 1 \,.
\ee

The curve $\crv$ lies in the ambient space $\as$. In order for it to lie on the double-cover $\cyu$ of the orientifold $\cyo$, the embedding map must satisfy
\be
\Xi^2(u,v) = h\big(X(u,v)\big) \,,
\ee
for all $[u:v]$. This is a degree $2\degr{\crv} $ equation in $[u:v]$, and so gives $2\degr{\crv} +1$ constraints on the parameters of the embedding map. This is precisely the number of parameters for the embedding, so the embedded curves are rigid.

For a given curve $\crv$, we wish to count the number of constraints placed on the D7-brane equation upon demanding that the D7-brane contain this curve. Demanding that the D7-brane inside $\cyu$ contain a specific curve, $\crv$, corresponds to imposing the equation $\eta(x)^2 =\xi^2 \chi(x) $ of the D7-brane for the entire image of the curve embedding,
\be
\eta\big(X(u,v))^2 - \Xi^2 (u,v) \chi\big(X(u,v)\big) = 0 \,.
\ee
As a degree $8\degr{\crv} $ equation in $[u:v]$, this gives $8\degr{\crv} +1$ constraints on the coefficients in the polynomials $\eta(x)$ and $\chi(x)$. Hence, the number of constraints on the D7-brane is
\be
\stbmod{\crv} = 8\degr{\crv} +1 \,.
\label{eq:mod_stab_tor}
\ee
We note that this result agrees with the general computation in Equation~\eqref{eq:mod_stab_gen}, when that expression is restricted to a genus-zero curve ($g_\crv=0$), and that it also agrees with the calculation for a $\mbb{P}^3$ base  in Equation~\eqref{eq:mod_stab_p3}.

\subsubsection*{Resolving the D7-brane and computing the tadpole contribution}

In Appendix~\ref{app:tadpole} we have computed the D3 charge of the stabilizing D7 worldvolume flux for a general compactification. This computation requires one to resolve the D7-brane, and the argumentation we used was rather abstract. Here we give a more explicit description of the resolution of the D7-brane when the base is toric.

The resolution is performed by blowing-up the ambient space $\as$ of the double-cover $\cyu$ along the singular locus of the D7-brane, $\db$, giving rise to a new auxiliary ambient space $\asl$ which contains a smooth surface $\dbl$ which is the resolved D7-brane.

We construct the auxiliary ambient space as follows. The singular locus of the D7-brane $\db$ occurs at $\xi=\eta(x)=0$. We blow up this locus by introducing the homogeneous coordinates $[s:t]$ of a $\mbb{P}^1$, and introducing an additional equation
\be
t \xi = s \eta(x) \,.
\ee
This gives rise to a (singular) toric variety $\asl$ with weight system
\be
\asl \sim
\begin{array}{c c c c c c}
x_0 & \ldots & x_n & \xi & s & t \\
\hline
q_1^0 & \cdots & q_1^n & \sum_j q_1^j & 0 & 3\sum_j q_1^j \\
\vdots & \ddots & \vdots & \vdots & \vdots & \vdots  \\
q_m^0 & \cdots & q_m^n & \sum_j q_m^j & 0 & 3\sum_j q_m^j  \\
0 & \cdots & 0 & 0 & 1 & 1 
\end{array} ~.
\ee
Associated to each row of the weight system is a divisor class, which we write as $\{\overline{H}_1 \,, \ldots \,, \overline{H}_m \,, [s]\}$. We also define $[-\candl] \equiv \sum_i \overline{H}_i$, which is the lift to $\asl$ of the anti-canonical divisor $-\cand{\fthb}$ of the F-theory base, $\fthb$. We note for later use the divisor class equivalence $[t] \sim [s]+3[-\candl]$, and the intersection property $[t] \cdot [s] = 0$, or equivalently
\be
[s] \cdot \big([s] + 3[-\candl]\big) = 0 \,.
\ee

In the blow-up, the surface $\db$ wrapped by the D7-brane is lifted to a non-singular surface, $\dbl$, inside $\asl$, which is given by the complete intersection
\be
\dbl: ~~ \xi t = \eta(x) s ~\cap~ t^2 = \chi(x) s^2 ~\cap~ \xi^2 = h(x) \,,
\ee
and which is hence described by the divisor class intersection
\be
[\dbl] = \big(4[-\candl]+[s]\big) \cdot \big(6[-\candl]+2[s]\big) \cdot \big(2[-\candl]\big) \,.
\ee
We can also compute the (object that restricts from $\asl$ to $\dbl$ to give the) Chern class of the $\dbl$ by the adjunction formula:
\be
c\big(\asl,\tanb{\dbl}\big) = \frac{c\big(\asl,\tanb{\asl}\big)}{c\big(\asl,\nrmb{\dbl}{\asl}\big)} = \frac{1+\big(5[-\candl]+2[s]\big)+\ldots}{1+\big(12[-\candl]+3[s]\big)+\ldots} = 1+\big(-7[-\candl]+[s]\big)+\ldots \,,
\label{eq:ch_auxsurf}
\ee
where we have only written out the first Chern classes in each term.

Finally, one can hence compute a lower bound for the contribution to the D3 tadpole of the stabilizing worldvolume flux: 
\be
\begin{aligned}
Q_\crv = -\frac{1}{4}(\crvl-\crvl')^2 \geq -\frac{1}{2}\,\crvl^2 &= \frac{1}{2}\big(\chi(\crvl)- c_1(\tanb{\dbl}) \cdot \crvl\big) \\
&= (1-g_\crv) + \frac{7}{2} (-\candl) \cdot \crvl  \equiv (1-g_\crv) + \frac{7}{2} \degr{\crv}
\,,
\end{aligned}
\label{eq:tadp_cont_tor}
\ee
where the intersection in the last line is taken on $\asl$, and where we have again defined the canonical degree $\degr{\crv} \equiv (-\candl) \cdot \crvl$. This agrees with the general result in Equation~\eqref{eq:general_q}.

Hence, the explicit calculations of the number of stabilized moduli and of the tadpole induced by fluxes in a compactification with a toric base reproduce the results in Section~\ref{sec:gen} for a more general compactification with a weak Fano base.
The examples worked out in this Appendix correspond to the vertical $g_\crv=0$ line in \fref{fig:constraint_plot}.

\providecommand{\href}[2]{#2}\begingroup\raggedright\endgroup

\end{document}